\documentclass[preprint]{elsarticle}

\usepackage{lineno,hyperref}
\usepackage{amsmath,amssymb,xfrac}
\usepackage{amsfonts}
\usepackage{graphicx,subcaption,float}
\usepackage{epstopdf}
\usepackage{tikz}
\usepackage{wasysym}
\usepackage{tabularx}
\usepackage{ragged2e}
\newcolumntype{Y}{>{\RaggedRight\arraybackslash}X} 
\usepackage{booktabs}

\modulolinenumbers[200]

\journal{Journal of Computational Physics}

\begin{document}

\begin{frontmatter}

\title{An implicit ODE-based numerical solver for the simulation of the Heisenberg-Euler equations in 3+1 dimensions }

\author[LMU,MaxPlanck]{Arnau Pons Domenech}
\ead{arnau.pons@physik.uni-muenchen.de}
\author[LMU]{Hartmut Ruhl}
\ead{hartmut.ruhl@physik.uni-muenchen.de}
\address[LMU]{Ludwig-Maximilians-Universit\"at M\"unchen\\Theresienstr.37, 80333 M\"unchen, Germany}
\address[MaxPlanck]{Max-Planck-Institut f\"ur Quantenoptik\\Hans-Kopfermann-Str.1, 85748 Garching, Germany}

\begin{abstract}
With the advent of ultra-high power lasers the nonlinear nature of the
vacuum of quantum electrodynamics (QED) can be probed. Due to the
highly nonlinear structure of the underlying equations new numerical
algorithms are required. A numerical scheme for simulating the
nonlinear optical effects of the QED vacuum in up to 3 spatial
dimensions plus time is derived. Its properties are discussed. The
validity of the numerical approach is verified with the help of known
analytic results. The algorithm is used to explore nonlinear all optical effects
of the nonlinear vacuum for which analytic methods are inefficient or
impossible.
\end{abstract}


\end{frontmatter}

\linenumbers

\section{Introduction \label{Sec:Introduction}}
It has long been suspected that the structure of the QED vacuum can lead
to nonlinear polarization and magnetization terms in the vacuum wave
equations for electromagnetic (em)-fields. First derived by Heisenberg
and Euler \cite{heisenberg1935original} and later introduced into the
QED framework by Schwinger \cite{schwinger} the breaking of the
spatial isotropy of the vacuum by strong em-fields is expected to lead
to many new effects such as vacuum birefringence \cite{baier1967vacuumIndexes},
diffraction \cite{Tommasini2010diffraction,Monden2011apertureDiffraction},
scattering \cite{king2012scattering}, dispersion
\cite{adler1971splitAndDisperse}, reflection
\cite{Gies2013reflection}, the generation of higher harmonics
\cite{Boehl1,Boehl2} and others. While the theory has long been known
the theoretical investigation of these effects has started only
recently. None of them has yet been measured. With the advent of
Petawatt (PW)-class lasers (e.g. ELI-NP and others
\cite{danson2015petawattclass,Cowan2013HIBEF,miyanaga2006FIREX}) it is
expected that some of the nonlinear vacuum effects might become
accessible experimentally in the near future.

While there are by now many analytical endeavors to compute nonlinear
vacuum effects \cite{Boehl1,Boehl2,Gies1,Greger1,Greger2} there are
only a few on the numerical side \cite{Boehl1}. Analytical calculations have many
limitations such as low energies, the requirement of special pulse
shapes and many more. They become highly complex when attempting to
reproduce realistic experimental settings. Furthermore, the analytical theories can
only consider light by light interaction while, e.g., plasma
effects have to be neglected.

Thus, in the present paper a numerical algorithm is introduced that
only makes use of a minimal set of assumptions and has the power to
augment analytical investigations. With the numerical approach
presented in this paper it is possible explore almost all nonlinear
effects of the quantum vacuum simultaneously and not in isolation and
for any given setting. The algorithm is efficient and accurate and can
be used in up to three spatial dimensions plus time ($3+1$). While the
algorithm in \cite{Boehl1} works in $1+1$ the extension to $3+1$ is
nontrivial as is shown in section \ref{Sec:Numerical}.

The paper is divided into 8 sections. In section \ref{Sec:LHE} the
Heisenberg-Euler Lagrangian and its weak-field expansion are
introduced. The weak-field expansion is then used in section
\ref{Sec:Maxwell} to derive a modified set of Maxwell equations, that
account for light by light scattering. In section \ref{Sec:Numerical}
the derivation of the numerical algorithm based on the
Heisenberg-Euler (HE) equations in weak-field expansion is derived. In
section \ref{Sec:Properties} the properties of the numerical algorithm are
discussed. In section \ref{Sec:Analytical} the most prominent
nonlinear QED vacuum effects are reviewed \cite{Boehl1,Gies1,Greger1} 
and in section \ref{Sec:Results} they are benchmarked with simulations
based on the numerical algorithm derived in the present
paper. Section \ref{Sec:Discussion} contains a summary of results and
their significance. Finally in section \ref{Sec:Outlook} an outlook
into possible future applications of the numerical algorithm is given.

\section{Heisenberg-Euler Lagrangian \label{Sec:LHE}}
It is useful to normalize the electromagnetic invariants $\mathcal{F},
\mathcal{G}$ and secular invariants $a,b$ to the critical field
strength given by $E_{cr}=m^2 c^3/e\hbar=1.3\times 10^{18}
\sfrac{\text{V}}{\text{m}}$ \cite{heisenberg1935original} leading to
\begin{align}\label{normalization_invariants}
\mathcal{F}=-\frac{F^{\mu\nu}F_{\mu\nu}}{4E_{cr}^2}=\frac{1}{2E_{cr}^2}\left(\vec{B}^2-\frac{\vec{E}^2}{c^2}\right),
&\qquad\mathcal{G}=-\frac{F^{\mu\nu}F^\ast_{\mu\nu}}{4E_{cr}^2}=\frac{\vec{E}\cdot\vec{B}}{c\,E_{cr}^2},\\[3pt]
	a=\sqrt{\sqrt{\mathcal{F}^2+\mathcal{G}^2}+\mathcal{F}},
&\qquad b=\sqrt{\sqrt{\mathcal{F}^2+\mathcal{G}^2}-\mathcal{F}},
\end{align}
where $F$ is the electromagnetic tensor and $F^\ast$ its dual. Using
these definitions the effective HE-Lagrangian representing the
interaction of a constant electromagnetic field with virtual
electron-positron pairs is given by \cite{schwinger}
\begin{align}\label{pureLHE}
\mathcal{L}_{HE}&=-\frac{m^4}{8 \pi ^2} \int_0^{\infty } \text{d}s \,
\frac{e^{-s}}{s^3} \left(\frac{s^2}{3}  \left(a^2-b^2\right)-1+a b
  s^2 \cot (a s) \coth (b s)\right) \, .
\end{align}
It has been shown both numerically
\cite{LheInNonConstantBackgroundNumerical} and analytically
\cite{LheInNonConstantBackgroundAnalytical} that the approximation of
a constant background em-field holds when the variations of the field
strength are on a much larger scale than the Compton wavelength
$\lambda_C=\tfrac{h}{m\,c}=2.426\times10^{-12}\text{m}$
\cite{ComptonWavelength}. 

\begin{figure}
	\centering
	\includegraphics[width=0.8\textwidth]{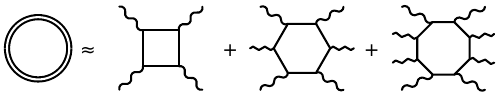}
	\caption{Depiction of the weak-field expansion of the
          HE-Lagrangian using Feynman diagrams. To the left the full
          seeded loop and on the right the 4, 6 and 8 photon box
          diagrams are shown. Note that these correspond to
          \eqref{LHE_T1}, \eqref{LHE_T2}, and \eqref{LHE_T3}
          respectively.
	\label{Fig:Diagramme}}
\end{figure}

The evaluation of the integral in \eqref{pureLHE} is
challenging. However, since the critical field is assumed to be much
larger than all other em-fields involved a weak-field approximation,
i.e., a Taylor series of the $\cot$ and $\coth$ functions for small
$a$ and $b$ is made. This yields the following weak field
approximation of the HE-Lagrangian
\begin{subequations}\label{approxLHE}
\begin{align}
\mathcal{L}_{HE}\approx &\quad\frac{m^4}{360 \pi ^2} \left(\,4\mathcal{F}^2+7\mathcal{G}^2\,\right)\label{LHE_T1}
	\\&+\frac{m^4}{630 \pi ^2} \left(8 \mathcal{F}^3+13\mathcal{F}\mathcal{G}^2\,\right)\label{LHE_T2}
	\\&+\frac{m^4}{945 \pi ^2} \left(48 \mathcal{F}^4+88 \mathcal{F}^2 \mathcal{G}^2+19 \mathcal{G}^4\,\right)\label{LHE_T3}
	\\&+\frac{4 m^4}{1485 \pi ^2} \left(160 \mathcal{F}^5+332
            \mathcal{F}^3 \mathcal{G}^2+127 \mathcal{F}
            \mathcal{G}^4\,\right).
\end{align}
\end{subequations}
Three things are worth noting: Firstly, \eqref{pureLHE} and
\eqref{approxLHE} only depend on $\mathcal{F}$ and
$\mathcal{G}$. Secondly, due to the definition of the
electromagnetic invariants in (\ref{normalization_invariants})
$\mathcal{F},\mathcal{G}\sim (E/E_{cr})^2$ holds. Therefore the term
\eqref{LHE_T1} is $\mathcal{O}((E/E_{cr})^4)$, term \eqref{LHE_T2}
$\mathcal{O}((E/E_{cr})^6)$ and so on. This implies that in the
weak-field regime, $E<E_{cr}$, the higher order terms become
negligible. The expansion in \eqref{approxLHE}
corresponds to processes with 4, 6, 8 and 10 photons
contributing to the closed loop as shown in figure \ref{Fig:Diagramme}.
Furthermore, computing the probabilitiy for pair production in a
constant electric field using the full HE-Lagrangian 
\cite{schwinger}
\begin{equation}\label{Eq:PairProd}
    	R_{e^+\;e^-}=\frac{\alpha
          E^2}{4\pi^3}\,\sum\limits_{n=1}^{\infty}\frac{1}{n^2}\,e^{-
          n\pi\, \frac{E_{cr}}{E}} \, .
\end{equation}
shows that the production of real pairs is exponentially suppressed.

\section{Modified Maxwell equations \label{Sec:Maxwell}}
For simplicity we use natural units $\hbar=c=1$ combined with
Lorentz-Heaviside units $e^2=4\pi\alpha$ . In these units the vacuum
Maxwell-Lagrangian is given by
 \begin{equation}
\mathcal{L}_{MW}=-\frac{F^{\mu\nu}F_{\mu\nu}}{4}\xrightarrow{E\rightarrow
  \frac{E}{E_{cr}}}-\frac{m^4}{4\pi\alpha}\frac{F^{\mu\nu}F_{\mu\nu}}{4}
\, .
\end{equation}
The Maxwell-Lagrangian combined with the HE-Lagrangian yields the
Lagrang\-ian for the propagation of light in vacuum
$\mathcal{L}=\mathcal{L}_{MW}+\mathcal{L}_{HE}$. Making use of the
Lagrange equations of the second kind and the cartesian representation
of the electromagnetic tensor results in
\begin{equation}\label{Eq:MaxPDE}
\partial_t \left(\vec{E}+\mu\frac{\partial \mathcal{L}_{HE}}{\partial
    \vec{E}}\right)=\nabla\times\left(\vec{B}-\mu\frac{\partial
    \mathcal{L}_{HE}}{\partial \vec{B}}\right) \, ,
\end{equation}
where $\mu=4\pi\alpha/m^4$. With the help of \eqref{Eq:MaxPDE}
Maxwell's equations \cite{Maxwell} become
\begin{subequations}\label{MaxesEq}
\begin{align}
\partial_t \vec{B} &= -\nabla\times\vec{E} \, , \label{MaxesEqB}\\
\partial_t
  \left(\vec{E}+\vec{P}\right)&=\nabla\times\left(\vec{B}-\vec{M}\right)
                                \, , \label{MaxesEqE}
\end{align}
\end{subequations}
where
\begin{equation}\label{MPtoLHE}
	\vec{P}=\mu\frac{\partial \mathcal{L}_{HE}}{\partial
          \vec{E}}\quad\text{and}\quad\vec{M}=\mu\frac{\partial
          \mathcal{L}_{HE}}{\partial \vec{B}} \, .
\end{equation}
Equations (\ref{MaxesEqB}) and (\ref{MaxesEqE}) from a set of
nonlinear dynamical equations for the em-fields. They will be solved
numerically in the present paper.

\section{Numerical Method \label{Sec:Numerical}}
\subsection{Linear case}
The linear case is considered first. Assuming $\vec{P}=\vec{M}=0$
we find, e.g., for the curl of $\vec B$ in (\ref{MaxesEqE})
\begin{align}
&\nabla \times 
\begin{pmatrix}
 B_x \\ B_y\\B_z 
\end{pmatrix}
=
\begin{pmatrix}
\partial_y B_z - \partial_z B_y\\
\partial_z B_x - \partial_x B_z\\
\partial_x B_y - \partial_y B_x 
\end{pmatrix}
\\&=
\underbrace{\begin{pmatrix}
0&0&0\\
0&0&-1\\
0&1&0 
\end{pmatrix}}_{\mathbf{Q_x}}\partial_x\vec{B}
+
\underbrace{\begin{pmatrix}
0&0&1\\
0&0&0\\
-1&0&0 
\end{pmatrix}}_{\mathbf{Q_y}}\partial_y\vec{B}
+
\underbrace{\begin{pmatrix}
0&-1&0\\
1&0&0\\
0&0&0 
\end{pmatrix}}_{\mathbf{Q_z}}\partial_z\vec{B} \nonumber\\
&=\sum_{i\in \{x,y,z\}}\mathbf{Q}_i\partial_i \vec{B} \nonumber \, . 
\end{align}
where $\mathbf{Q}_i=(\epsilon_{ijk})_{j,k}\in\mathbb{R}^{3\times 3}$
implying that equations \eqref{MaxesEq} can be written as
\begin{equation} \label{LinPropEq}
\partial_t\vec{f}=\sum_{i\in\{x,y,z\}}\begin{pmatrix}0
&\mathbf{Q}_i\\-\mathbf{Q}_i&0\end{pmatrix}\partial_i\vec{f} \, ,
\end{equation}
where $\vec{f}=( \vec{E}, \vec{B})^T\in\mathbb{R}^6$. 

In order to convert \eqref{LinPropEq} into a finite set of
ordinary differential equations (ODEs) the derivatives in space are
replaced by a finite difference scheme. To do this a grid of
$N=N_x\cdot N_y\cdot N_z$ equally spaced points
$\mathcal{B}=\{x_{a,b,c}=(a\Delta_x,b\Delta_y,c\Delta_z)\}\subset\mathbb{R}^3$
is introduced and the vector $\vec{f}$ of the em-field at the grid points
$\vec{f}_{a,b,c}=\vec{f}(x_{a,b,c})$ is merged into a new vector
$\vec{F}\in\mathbb{R}^{6N}$, such that $(\vec{f}_{a,b,c})_l=F_{\mathcal{I}(a,b,c)+l}$. Here
$\mathcal{I}:[0,N_x]\times[0,N_y]\times[0,N_z]\rightarrow[0,6N]$ is a
bijection, which assigns each set of coordinates
$(a,b,c)\in\mathbb{N}^3$ a position $\mathcal{I}(a,b,c)\in\mathbb{N}$
in the data vector.

The derivative of the field can now be approximated by a weighted
difference of the values of $\vec{F}$. To illustrate this, w.l.o.g.,
the derivative in the $x$ direction in \eqref{LinPropEq} is considered
\begin{equation}
\partial_x \vec{f}_{a,b,c}=\sum_\nu \mathcal{S}_\nu 
\vec{f}_{a+\nu,b,c}+\mathcal{O}(\Delta_x^n) \, , \label{discrete_derivative_x}
\end{equation}
where the order of accuracy $n$ is given by the choice of the stencil
$\widehat{\mathcal{S}}$ with elements $\mathcal{S}_n$. The size of the
stencil is defined as 
\begin{equation}
\big|\widehat{\mathcal{S}}\big|=\big|\{n\in\mathbb{Z} \, | \,
\mathcal{S}_n\neq0\}\big| \, .
\end{equation}
Upwind differencing of the 1st derivative of an arbitrary function
$g$ to 2nd order implies \citep{NumericalRecipies} 
\begin{equation}
\frac{\partial g(x)}{\partial x}=\frac{g(x+\Delta_x)-g(x)}{\Delta_x}
\, .
\end{equation}
The corresponding stencil is defined as
\begin{equation}
\mathcal{S}_0=\mathbf{1}\,\left(-\frac{1}{\Delta_x}\right),\quad
\mathcal{S}_1=\mathbf{1}\,\frac{1}{\Delta_x}, \quad \mathcal{S}_i=0
\quad \forall i \notin \{0,1\}
\label{Eq:upStencil}
\end{equation}
and has the size $\big|\widehat{\mathcal{S}}\big| = 2$. In the case
of the downwind differencing
\begin{equation}
\frac{\partial g(x)}{\partial x}=\frac{g(x)-g(x-\Delta_x)}{\Delta_x}
\end{equation}
the corresponding stencil is defined as
\begin{equation}
\mathcal{S}_{-1}=\mathbf{1}\,\left(-\frac{1}{\Delta_x}\right),\quad
\mathcal{S}_0=\mathbf{1}\,\frac{1}{\Delta_x}, \quad \mathcal{S}_i=0
\quad \forall i \notin \{-1,0\} \, . \label{Eq:downStencil}
\end{equation}

There are two things to be considered when choosing
$\widehat{\mathcal{S}}$. Firstly, with larger size of the stencil
both, the accuracy of the numerical scheme and the numerical load of
the algorithm increase. For the solver presented in this paper
4th order differences are used. The second consideration to be made is
biasing. Choosing a symmetric distribution
$\mathcal{S}_{-k}=\mathcal{S}_k$ of stencil points leads to
$\mathcal{S}_0=0$ causing two disconnected grids. Disconnected grids
cause mesh drift instabilities \cite{NumericalRecipies}. Avoiding the latter by
biasing the stencil, i.e., by taking an asymmetric stencil results in
an anisotropic space, which reduces the propagation speed of em-waves
in the direction towards which the stencil is heavier weighted.

While the mesh drifting cannot be avoided for symmetric stencils
propagation speed errors due to biased stencils however can.
To see this the expression 
\begin{equation}\label{Eq:DiagQ}
\begin{pmatrix}0&\mathbf{Q}_x\\-\mathbf{Q}_x&0\end{pmatrix}=\mathbf{R}_x^T 
\text{diag}(0,1,1,0,-1,-1) \mathbf{R}_x 
\end{equation}
in \eqref{LinPropEq} is first diagonalized, where 
\begin{equation}
\mathbf{R}_x=\frac{1}{\sqrt{2}}\begin{pmatrix}
 0& 0& 0& 0& 0& 0\\
 0& 1& 0& 0& 0&-1\\
 0& 0& 1& 0& 1& 0\\
 0& 0& 0& 0& 0& 0\\
 0& 1& 0& 0& 0& 1\\
 0& 0& 1& 0&-1& 0 
\end{pmatrix} 
\end{equation}
leading with the help of (\ref{discrete_derivative_x}) to
\begin{equation}
\begin{pmatrix}0&\mathbf{Q}_x\\-\mathbf{Q}_x&0\end{pmatrix} \, 
\partial_x \vec f
\approx \begin{pmatrix}0&\mathbf{Q}_x\\-\mathbf{Q}_x&0\end{pmatrix} \, 
\mathcal{D}_x \vec{f}_{a,b,c} \, ,
\end{equation}
where
\begin{equation}\label{Eq:RotatedFiniteDerivative}
\mathcal{D}_x \vec{f}_{a,b,c} = \sum_\nu \mathbf{R}_x^T
                                 \mathcal{S}_\nu
                                 \mathbf{R}_x\vec{f}_{a+\nu,b,c} \, ,
\quad
\mathbf{R}_x \vec f = \frac{1}{\sqrt{2}} \, 
\left(
\begin{array}{c}
0\\
E_y - B_z\\
E_x + B_y\\
0\\
E_y + B_z\\
E_z - B_y\\
\end{array}
\right) \, . 
\end{equation}
Equations \eqref{Eq:DiagQ} and \eqref{Eq:RotatedFiniteDerivative} imply that in the
rotated frame the solution of (\ref{LinPropEq}) propagates only in the
direction, which is given by the sign of the eigenvalues of the diagonal
matrix in equation \eqref{Eq:DiagQ}. This implies that $E_y-B_z$ and
$E_z+B_y$ both propagate along the positive $x$-direction while
$E_y+B_z$ and $E_z-B_y$ propagate along the negative $x$-direction.
Choosing downwind biasing for fields propagating along the positive
$x$-axis, upwind biasing for fields propagating along the negative
$x$-axis and rotating back afterwards suppresses the propagation speed
error. As a consequence no mesh drifting and no propagation speed 
asymmetry are expected.

Making use of the up- and downwind stencils to 2nd order as defined in
\eqref{Eq:upStencil} and \eqref{Eq:downStencil} for $\mathbf{R}_x \vec
f$ for the different directions of em-wave propagation results in the
following matrices for the stencil elements
\begin{align} \label{Eq:upDownStencil}
\mathcal{S}_{-1} &=\text{diag}(0,-1,-1,0,0,0)\,\frac{1}{\Delta_x} \, , \\
\mathcal{S}_0 &=\text{diag}(0,1,1,0,-1,-1)\,\frac{1}{\Delta_x} \, , \nonumber \\
\mathcal{S}_{1} &=\text{diag}(0,0,0,0,1,1)\,\frac{1}{\Delta_x} \, , \nonumber \\
\mathcal{S}_i &=0 \quad \forall i \notin \{-1,0,1\}  \, . \nonumber
\end{align}
A compact notation for \eqref{Eq:upDownStencil} is
\begin{align}
\mathcal{S}_{\nu} &=\text{diag} \left( 0, s^+_{\nu}, s^+_{\nu}, 0,
                   s^-_\nu, s^-_\nu \right)\,\frac{1}{\Delta_x} \, , \\
	s^+_\nu\big|_{\nu=-1,\ldots,1}&=\left\{-1, 1, 0 \right\} \,
                                        \nonumber \\
	s^-_\nu\big|_{\nu=-1,\ldots,1}&=\left\{0, -1, 1 \right\} \,\nonumber \\
	s^{+/-}_\nu&=0\;\quad\forall \nu:|\nu|>1 \, . \nonumber 
\end{align}
The stencils used in the simulations in the present paper are
accurate up to the 4th order, where the
values of the stencils at the diagonals $s^+_\nu$ and $s^-_\nu$ in the
forward $+$ and backward $-$ directions are given by
\begin{align}
	s^+_\nu\big|_{\nu=-3,\ldots,3}&=\left\{-\frac{1}{12}\;,\hphantom{+}\frac{1}{2}\;,
-\frac{3}{2}\;,\hphantom{+}\frac{5}{6}\;,\hphantom{+}\frac{1}{4}\;,\hphantom{+}0\;,
\hphantom{+}0\right\} \, , \\
	s^-_\nu\big|_{\nu=-3,\ldots,3}&=\left\{\hphantom{+}0\;,\hphantom{+}0\;,-\frac{1}{4}\;,
-\frac{5}{6}\;,\hphantom{+}\frac{3}{2}\;,-\frac{1}{2}\;,\hphantom{+}\frac{1}{12}\right\}
                                        \, ,\nonumber\\
	s^{+/-}_\nu&=0\;\quad\forall \nu:|\nu|>3 \, . \nonumber 
\end{align}
Note that the rotation matrices will always sort the eigenvalues of
adiag $\left( -\mathbf{Q}_i, \mathbf{Q}_i \right)$, where $i=x,y,z$
in the same way and thus the $\mathcal{S}_\nu$ is identical in all three 
spatial directions.

\subsection{Nonlinear case}
Using the same approach for equations \eqref{MaxesEqB} and \eqref{MaxesEqE} yields
\begin{equation}\label{PropEquation}
\partial_t\vec{f}=\left( \mathbf{1}+\mathbf{A} \right)^{-1}\sum_i
\mathbf{B}_i \, \partial_i\vec{f} \, ,
\end{equation}
where
\begin{align}
\mathbf{A}&=\begin{pmatrix}\mathbf{J}_{\vec{P}}(\vec{E})&\mathbf{J}_{\vec{P}}(\vec{B})\\0&0\end{pmatrix}
  \, , \\
\mathbf{B}_i&=\begin{pmatrix}-\mathbf{Q}_i\mathbf{J}_{\vec{M}}(\vec{E})&\mathbf{Q}_i\left(\mathbf{1}
-\mathbf{J}_{\vec{M}}(\vec{B})\right)\\-\mathbf{Q}_i&0\end{pmatrix}
\end{align}
and $\mathbf{J}_{\vec{A}}(\vec{B})$ is the Jacobi matrix of $\vec{A}$ with respect to $\vec{B}$. 
The problem now is that the eigenvectors of the $\mathbf{B}_i$ are no
longer constant since they depend on the em-fields. This implies that
the matrices $\mathbf{R}_i$ with $i=x,y,z$ depend on the em-field
configuration and cannot be easily calculated analytically. However, the
nonlinear corrections due to $\mathbf{A}$ and $\mathbf{B}$ in
\eqref{PropEquation} are small. As a consequence the effect of the
nonlinear contributions in \eqref{PropEquation} on the eigenvalues and
eigenvectors of the matrix $\mathbf{B}$ can be neglected. Thus, the same
rotations used for the derivatives in the linear case can be used and the
numerical artifacts due to biasing are still suppressed in the
nonlinear case. By the same rationale the matrix inversion in
\eqref{PropEquation} can be performed by using the geometrical series
$(\mathbf{1}+\mathbf{A})^{-1}=\mathbf{1}-\mathbf{A}+\mathbf{A}^2-\mathcal{O}(\mathbf{A}^3)$
and neglecting higher order terms. This approximation is justified since
contributions from the higher order terms fall below the desired accuracy.

The nonlinear scheme is
\begin{align}\label{PropEquationNonlin}
\partial_t\vec{f}&=\left( \mathbf{1} - \mathbf{A} \right) \sum_i
\mathbf{B}_i \, \mathcal{D}_i\vec{f} \, , \\
\mathcal{D}_x \vec f &= \sum_\nu \mathbf{R}^T_x \mathcal{S}_\nu 
                       \mathbf{R}_x \, \vec f_{a+\nu, b, c} \, , \\
\mathcal{D}_y \vec f &= \sum_\nu \mathbf{R}^T_y \mathcal{S}_\nu 
                       \mathbf{R}_y \, \vec f_{a, b+\nu, c} \, , \\
\mathcal{D}_z \vec f &= \sum_\nu \mathbf{R}^T_z \mathcal{S}_\nu 
                       \mathbf{R}_z \, \vec f_{a, b, c+\nu} \, ,
\end{align}
where
{\small
\begin{equation}
\mathbf{R}_x=\frac{1}{\sqrt{2}}\begin{pmatrix}
 \sqrt{2}& 0& 0& 0& 0& 0\\
 0& 1& 0& 0& 0&-1\\
 0& 0& 1& 0& 1& 0\\
 0& 0& 0& \sqrt{2} & 0& 0\\
 0& 1& 0& 0& 0& 1\\
 0& 0& 1& 0&-1& 0 
\end{pmatrix} \, , \quad
\mathbf{R}_x \vec f = \frac{1}{\sqrt{2}} \, 
\left(
\begin{array}{c}
\sqrt{2} \, E_x\\
E_y - B_z\\
E_x + B_y\\
\sqrt{2} \, B_x\\
E_y + B_z\\
E_z - B_y\\
\end{array}
\right)
\end{equation}}
and
{\small
\begin{equation}
\mathbf{R}_y=\frac{1}{\sqrt{2}}\begin{pmatrix}
 0 &\sqrt{2}& 0& 0& 0& 0\\
 -1 & 0& 0& 0& 0&-1\\
 0& 0& -1& 1& 0& 0\\
 0& 0& 0& 0 & \sqrt{2} & 0\\
 -1 & 0& 0& 0& 0& 1\\
 0& 0& -1& -1& 0 & 0 
\end{pmatrix} \, , \quad
\mathbf{R}_x \vec f = \frac{1}{\sqrt{2}} \, 
\left(
\begin{array}{c}
\sqrt{2} \, E_y\\
- B_z - E_x\\
B_x - E_z\\
\sqrt{2} \, B_y\\
B_z - E_x\\
- B_x - E_z\\
\end{array}
\right) 
\end{equation}}
and
{\small
\begin{equation}
\mathbf{R}_z=\frac{1}{\sqrt{2}}\begin{pmatrix}
 0 & 0& \sqrt{2}& 0& 0& 0\\
 1& 0& 0& 0& -1&0\\
 0& 1& 0& 1& 0& 0\\
 0& 0& 0& 0 & 0& \sqrt{2}\\
 1& 0& 0& 0& 1& 0\\
 0& 1& 0& -1&0& 0 
\end{pmatrix} \, , \quad
\mathbf{R}_z \vec f = \frac{1}{\sqrt{2}} \, 
\left(
\begin{array}{c}
\sqrt{2} \, E_z\\
-B_y + E_x\\
B_x + E_y\\
\sqrt{2} \, B_z\\
B_y + E_x\\
-B_x + E_y\\
\end{array}
\right) \, . 
\end{equation}}
Due to the nonlinear contributions coupling to the derivative in
propagation direction the stencils become
\begin{align}
\mathcal{S}_{\nu} &=\text{diag} \left( s^+_\nu, s^+_{\nu}, s^+_{\nu}, s^-_\nu,
                   s^-_\nu, s^-_\nu \right)\,\frac{1}{\Delta} \, ,
\end{align}
where $\Delta$ are the spatial resolutions along the directions $x, y, z$.

\subsection{Solving the ODE and processing the data}
The ODE for the propagation of light defined by \eqref{PropEquation}
can now be solved by any generic ODE solver. In this paper the {\it
  CVODE} solver from the {\it SUNDIALS} bundle is used
\cite{SUNDIALS}. The post-processing of the simulation data for this
paper has been performed with {\it Mathematica} \cite{Mathematica}.

\subsection{Parallelisation of the algorithm \label{Sec:paralelisationTheory}}
The numerical method presented in this paper can be parallelized. In
order to distribute the computational load over multiple processing
cores the lattice is sliced into smaller sub-lattices. The
communication between compute nodes is limited to the exchange of
boundary values for the computation of the discrete space
derivatives in \eqref{PropEquationNonlin}. The computation of the right hand side of
\eqref{PropEquationNonlin} and the subsequent time step update are then a
purely local problem.
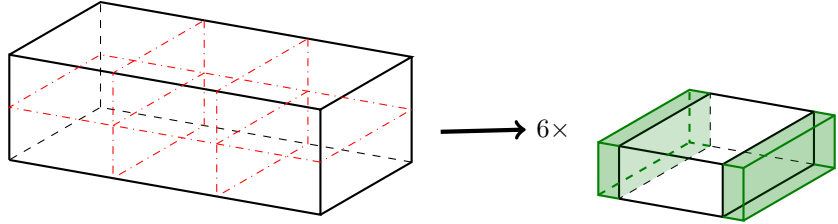
\begin{figure}[h]
\begin{center}
\begin{tikzpicture}[scale=0.7,x={(-10:0.2cm)},y={(90:1cm)},z={(210:1cm)}]
	\draw [thick](0,0,0)--(30,0,0)--(30,2,0)--(0,2,0)--(0,0,0);
	\draw [thick](0,2,0)--(0,2,-2)--(30,2,-2)--(30,2,0);
	\draw [thick](30,2,-2)--(30,0,-2)--(30,0,0);
	\draw [dashed](30,0,-2)--(0,0,-2)--(0,0,0);
	\draw [dashed](0,0,-2)--(0,2,-2);
	\draw [dashdotted,red](10,2,-2)--(10,0,-2)--(10,0,0)--(10,2,0)--(10,2,-2);
	\draw [dashdotted,red](20,2,-2)--(20,0,-2)--(20,0,0)--(20,2,0)--(20,2,-2);
	\draw [dashdotted,red](30,1,0)--(0,1,0)--(0,1,-2)--(30,1,-2)--(30,1,0);
	\draw [dashdotted,red](10,1,0)--(10,1,-2);
	\draw [dashdotted,red](20,1,-2)--(20,1,0);
	\draw [ultra thick,->] (35,1,-1.5)--(41,1,-2)node[right]{$6\times$};
	\draw [thick](50,0,-2)--(60,0,-2)--(60,1,-2)--(50,1,-2)--(50,0,-2);
	\draw [thick](50,1,-2)--(50,1,-4)--(60,1,-4)--(60,1,-2);
	\draw [thick](60,1,-4)--(60,0,-4)--(60,0,-2);
	\draw [dashed](60,0,-4)--(50,0,-4)--(50,0,-2);
	\draw [dashed](50,0,-4)--(50,1,-4);
	\draw [thick,black!50!green](60,0,-2)--(62,0,-2)--(62,1,-2)--(60,1,-2);
	\draw [thick,black!50!green](60,0,-4)--(62,0,-4)--(62,1,-4)--(60,1,-4);
	\draw [thick,black!50!green](62,0,-2)--(62,0,-4);
	\draw [thick,black!50!green](62,1,-2)--(62,1,-4);
	\draw [thick,black!50!green](50,0,-2)--(48,0,-2)--(48,1,-2)--(50,1,-2);
	\draw [thick,black!50!green,dashed](50,0,-4)--(48,0,-4)--(48,1,-4);
	\draw [thick,black!50!green,dashed](48,0,-2)--(48,0,-4);
	\draw [thick,black!50!green](48,1,-2)--(48,1,-4)--(50,1,-4);
	\fill [black!50!green,fill opacity =0.3] (62,0,-2)--(62,1,-2)--(62,1,-4)--(62,0,-4)--cycle;
	\fill [black!50!green,fill opacity =0.3] (60,1,-2)--(62,1,-2)--(62,1,-4)--(60,1,-4)--cycle;
	\fill [black!50!green,fill opacity =0.3] (62,0,-2)--(62,1,-2)--(60,1,-2)--(60,0,-2)--cycle;
	\fill [black!50!green,fill opacity =0.2] (50,0,-2)--(50,1,-2)--(50,1,-4)--(50,0,-4)--cycle;
	\fill [black!50!green,fill opacity =0.3] (48,1,-2)--(50,1,-2)--(50,1,-4)--(48,1,-4)--cycle;
	\fill [black!50!green,fill opacity =0.3] (50,0,-2)--(50,1,-2)--(48,1,-2)--(48,0,-2)--cycle;
\end{tikzpicture}
\caption{Sketch of a lattice decomposition (left) into 6 lattice
  patches. The red dashdotted lines indicate the boundaries of
  the lattice patches. The green cuboids attached to the lattice patch (right)
  are the ghost layers that store the values from neighboring patches.
  For better visibility only the ghost layers in one direction are shown.\label{Fig:PatchSetup}}
\end{center}
\end{figure}

\section{Properties of the numerical scheme \label{Sec:Properties}}
\subsection{Propagation error and dispersion relation for the linear case}
The pulse propagation in the QED vacuum reduces to the usual
propagation in vacuum when only a single plane wave is
present. Furthermore, the propagation error scales as
$\mathcal{O}(\Delta^4$). As the derivative in time is solved using a
recursive implicit algorithm it can be assumed that the time
derivative is exact. When inserting the plane wave 
$$
\vec{E}(\vec{x},t)=\vec{\varepsilon}\,\text{e}^{-i(\omega
  t-\vec{k}\cdot\vec{r})} 
$$
in \eqref{PropEquation}, where $\vec \varepsilon$ is the polarization
vector of the latter, and solving for the dispersion relation it is obtained
\begin{equation}\label{Eq:Dispersion}
0=\text{det}\left(i\omega\mathbf{1}_6- \sum
  _{j\in\{x,y,z\}}\text{adiag}(\mathbf{Q}_j,-\mathbf{Q}_j)\mathbf{R}_j^\intercal
  \sum _{\nu} S_{\nu } e^{-i \nu  k_j \Delta_j} \mathbf{R}_j\right) \, .
\end{equation}
The $S_\nu$ are the values of the stencils as introduced in section
\ref{Sec:Numerical}. Note that the rotation matrices $\mathbf{R}_j$ in
\eqref{Eq:Dispersion} disappear. Equation \eqref{Eq:Dispersion} is
evaluated numerically. 

\begin{figure}[h]
\begin{subfigure}{.56\textwidth}
\includegraphics[width=\linewidth]{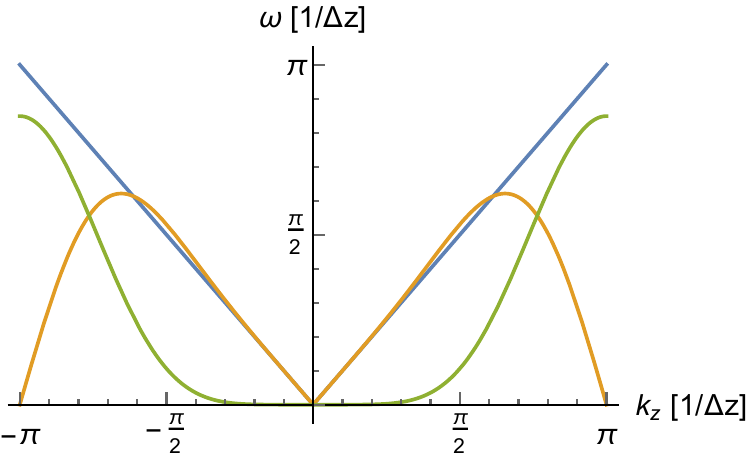}
\end{subfigure}
\begin{subfigure}{.43\textwidth}
\includegraphics[width=\linewidth]{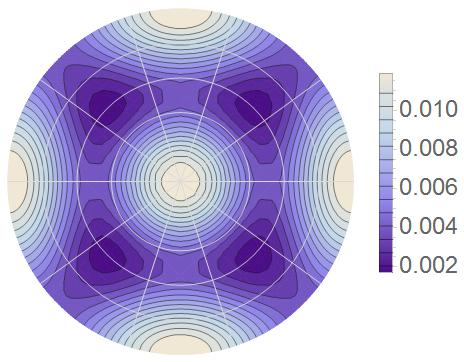}
\end{subfigure}
\caption{Left: $\omega$ vs. $k_z$ for a plane wave with $\vec{k}=(0\,
  0\, k_z)^T$. The blue line is the dispersion relation of the
  continuous vacuum. The orange one is the real part of $\omega$ as 
  computed in \eqref{Eq:Dispersion} and the green one
  is the imaginary part. Right: Spherical polar plot of the real part of 
  $\omega$ minus the $\omega$ in the continuous vacuum in 3D. The
  radial coordinate is $\theta$ with $0 \leq \theta < \pi$ and the
  angular variable is $\phi$ with $0 \leq \phi < 2\pi$. Note that
  $|\vec k|$ has been set to $0.8/\Delta z$. Larger values (brighter
  colors) imply that the wave propagates faster in that
  direction.\label{Fig:Dispersion}}
\end{figure}

The results are shown in Figure \ref{Fig:Dispersion}.
Figure \ref{Fig:Dispersion}(left) shows the frequency dependence of the
dispersion relation in $z$-direction. It can be seen that the
dispersion is well behaved for small $\Delta \, |\vec{k}| \ll 1$.
Furthermore, as $|\vec{k}|$ grows, i.e., the wavelength
becomes smaller, the phase velocity of the em-wave is above the light
speed. On top of that, the imaginary part of $\omega$, which causes
the lattice to absorb the electromagnetic field is negligible for
small $|\vec k|$ but becomes relevant at higher frequencies.

The dependence of the phase velocity on the propagation direction of
the wave is illustrated in figure \ref{Fig:Dispersion}(right). As can be
seen in the figure the wave propagates fastest along the
discretization directions $[1,0,0],[0,1,0]$ and $[0,0,1]$. This is an
expected behaviour for all Cartesian grid discretizations
\cite{NumericalRecipies,taflove2005computational}. The phase velocity
is lowest along the diagonal $[1,1,1]$. The difference between
$[1,1,1]$ and $[1,0,0]$ scales as $\mathcal{O}(|\vec{k}|^3)$ and
reaches a relative maximum of $1\%$ for frequencies below half of the
Nyquist frequency given by $f_{\text{Ny}}=\pi/2\Delta_z$.

\subsection{The computational load}
There are three key steps in the application of the algorithm: (i) The
computation of the derivatives, (ii) the computation on the right hand
side of \eqref{PropEquation}, and (iii) the integration of the ODE. 

While the number of operations involved in (i) for computing the
approximation of the space derivatives does not depend on the number
of lattice points considered, the derivatives have to be calculated
for each point separately and for each of the dimension considered
\begin{equation}
\mathcal{W}_\text{D}\propto D \cdot N_x \cdot N_y \cdot N_z \, ,\label{Eq:DerivationScalling}
\end{equation}
where $\mathcal{W}_\text{D}$ is the computational load.
A similar scaling holds for an integration step of
\eqref{PropEquation} addressed in (ii). The number of spatial
dimensions $D$ determines the number of summations. The right
hand side of \eqref{PropEquationNonlin} has to be evaluated for each
grid point of the lattice and thus the scaling becomes
\begin{equation}
\mathcal{W}_\text{M}\propto N_x \cdot N_y \cdot N_z \cdot(D+1) \, ,\label{Eq:MatrixScalling}
\end{equation}
where $\mathcal{W}_\text{M}$ is the computational load encountered for
the evaluation of \eqref{PropEquation}. We note that the term plus one in
\eqref{Eq:MatrixScalling} comes from the multiplication of the sum
with the $(1-A)$ matrix in \eqref{PropEquation}.

The load for the solution of the ODE (iii) depends on the choice of
the solver. For the {\it CVODE} solver used in the present paper
details can be found in \cite{SUNDIALS}. However, as the integration
of the ODE in \eqref{PropEquation} strongly depends on the problem
under consideration its scaling depends on the frequencies involved,
the strength of the nonlinearities, the precision required and so on
\begin{equation}
\mathcal{W}_\text{S}\propto \frac{1}{\Delta} \, N_x \cdot N_y \cdot
N_z \, ,
\end{equation}
where $\mathcal{W}_\text{S}$ is the computational load of the {\it
  CVODE} solver. Summarizing, an upper limit for the scaling of the
problem is given by
\begin{equation}
\mathcal{W} \propto \frac{1}{\Delta} \, N_x \cdot N_y \cdot N_z \cdot(D+1) .
\end{equation}

\subsection{Comparison to the Yee algorithm}
Comparison of the properties of the numerical scheme in the present
paper to the classical Yee solver for Maxwell's equations
\cite{taflove2005computational} shows several major
differences apart from the obvious increase in computational load since
the scheme in this paper is implicit and of 4th order accuracy while
the Yee solver is explicit and of 2nd order accuracy only. The major
differences are

\begin{enumerate}[i]
\item The computation of the nonlinearities in
  \eqref{PropEquationNonlin} does not require
  the extra effort of interpolating field values as 
  the scheme in this paper does not make use of 
  a staggered grid.
\item Since the numerical dispersion relation of the scheme in the
  present paper has an imaginary part aliasing modes are suppressed. 
  This is not the case for the Yee scheme.
\item The use of 4th order stencils in the scheme in this paper
  allows for 4th order accuracy of the present scheme while Yee is
  only 2nd order.
\item Through the use of propagation direction dependent biasing of
  the stencils in space a symmetric dispersion relation is achieved
  while mesh drifting is avoided. Staggering the grid as is done in
  the Yee scheme also avoids mesh drifting but causes loss of
  information needed for the computation of nonlinearities.
\end{enumerate}

\section{Benchmarling with analytical solutions \label{Sec:Analytical}}
The easiest test that can be performed to validate the accuracy and
efficiency of the numerical algorithm in the present paper is to check
it against the linear case. As $\mathcal{L}_{HE}$ vanishes for a
single planar laser pulse, the latter will propagate in the QED vacuum
as it would in the classical vacuum. The analytical solution for this
case is
\begin{equation}\label{Eq:AnaWave}
\vec{E}(\vec{x},t)=\vec{E}\left(\vec{x}-\frac{1}{3} \,
  \left(\frac{\omega}{k_x},\frac{\omega}{k_y},\frac{\omega}{k_z}\right)\,t,0\right)
\, ,
\end{equation}
where $\vec{k}$ is the wavenumber and $\omega/|\vec k|$ the phase
velocity. Equation \eqref{Eq:AnaWave} can be compared to simulation results to
determine the linear dispersion errors generated by the discretisation.

To validate the algorithm for the nonlinear part of the
vacuum wave equations numerical results are compared to those derived
analytically for some of the nontrivial effects of the nonlinear QED vacuum, a
comprehensive collection of which can be found in
\cite{king2015measuring}. 

In the present paper the flip of polarization and the generation of
higher harmonics are considered because they are two of the most well
know and best studied effects in literature. In addition, the main
contribution to polarization flips are due to 4-photon while the
asymptotic 2nd harmonics are due to 6-photon diagrams as depicted
in figure  \ref{Fig:Diagramme}. Hence, both 4- and 6-photon contributions
can be tested numeically.

\subsection{High harmonic generation}
\begin{figure}[h]
\centering
\includegraphics[scale=1.5]{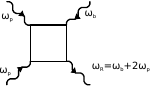}
\caption{Generation of higher harmonics due to 4-photon
  interaction. The two incoming photons to the left come from the
  probe $p$ and the incoming photon on the right comes from the
  strong background $b$. The frequency of the exiting photon has to be
  $\omega_R=\omega_b+2\omega_p$ due to energy conservation. \label{Fig:MergeBsp}}
\end{figure}

High harmonic generation is a direct consequence of energy
conservation. To see this the 4-photon scattering diagram in
figure  \ref{Fig:Diagramme} is considered. If the two legs to the left of
the diagram represent two incoming probe photons with $\omega_p$ and
the third leg represents the contribution of the pump photon
$\omega_b$ the frequency of the resulting photon has to be
$\omega_R=\omega_b+2\omega_p$ as is shown in figure  \ref{Fig:MergeBsp}.
As analytical estimates in 1D show 2nd harmonics generation is
expected to be several orders of magnitude smaller than polarization
flipping. However, it is expected to be more sensitive to the
pulse shapes involved in the interaction. In order to benchmark the
numerical results of 2nd harmonic generation the calculations
presented in \citep{Boehl1} are used for reference.

\subsection{Polarization flipping}
\begin{figure}[h]
\begin{center}
\begin{tikzpicture}[scale=0.8,x={(-10:0.2cm)},y={(90:1cm)},z={(210:1cm)}]
    \draw[blue] plot[domain=2:10,samples=200] (\x,{0.5*exp(-(\x-2*pi)*(\x-2*pi)/4)*cos(100*pi*(\x-2*pi))/sqrt(2)},{-0.5*exp(-(\x-2*pi)*(\x-2*pi)/4)*cos(100*pi*(\x-2*pi))/sqrt(2)});
    \draw[thick,blue,->] ({2*pi},0.4,-0.4)--({2*pi+1},0.4,-0.4) node [above]{$\vec{k}_{p}$} -- ({2*pi+2},0.4,-0.4);
    \draw[thick,orange] plot[domain=12:26,samples=200] (\x,{2*exp(-(\x-6*pi)*(\x-6*pi)/8)*cos(50*pi*(\x-6*pi))},0);
    \draw[thick,orange,->] ({6*pi},2.2,0)--({6*pi-1},2.2,0) node [above]{$\vec{k}_{b}$} -- ({6*pi-2},2.2,0);
    \draw[blue!50] plot[domain=28:36,samples=200] (\x,{0.5*exp(-(\x-10*pi)*(\x-10*pi)/4)*(cos(100*pi*(\x-10*pi))+0.3*sin(100*pi*(\x-10*pi)))/sqrt(2+0.1)},{-0.5*exp(-(\x-10*pi)*(\x-10*pi)/4)*(cos(100*pi*(\x-10*pi))-0.3*sin(100*pi*(\x-10*pi)))/sqrt(2+0.1)});
    \draw[thick,blue!50,->] ({10*pi},0.4,-0.4)--({10*pi+1},0.4,-0.4) node [above]{$\vec{k}'_{p}$} -- ({10*pi+2},0.4,-0.4);
    \draw [thick,->](0,-1.2,0) -- (0,2.2,0) node[above] {$E_y$};
    \draw [thick,->](0,0,-1.2) -- (0,0,1.7) node[left] {$E_z$};
    \draw [thick,gray,->](0,0,0) -- (0,1.5,0) node[left] {$\vec{\epsilon}_p$};
    \draw [thick,gray,->](0,0,0) -- (0,0.8,-0.8) node[right] {$\vec{\epsilon}_\parallel$};
    \draw [thick,gray,->](0,0,0) -- (0,0.8,0.8) node[left] {$\vec{\epsilon}_\perp$};
    \draw [thick,->](-5,0,0) -- (40,0,0)  node[right] {$x$};
\end{tikzpicture}
\caption{Qualitative sketch of the electric fields in a coaxial
  background-probe experiment for measuring vacuum birefringence. The
  probe (blue) travels through the counter-propagating background (orange)
  experiencing a polarisation flip due to the different refractive
  indices for different polarization directions
  \eqref{RefIndizes}. \label{Fig:BireSetup}}
\end{center}
\end{figure}
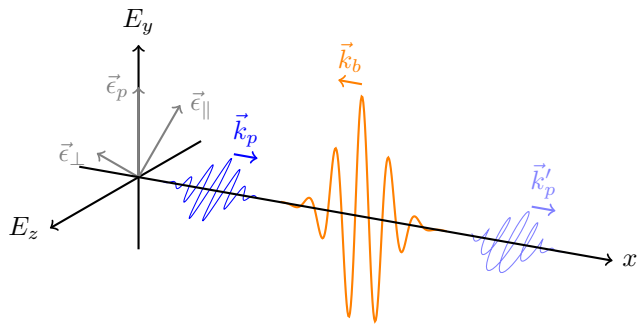
Polarization flipping is a result of vacuum birefringence. Here the
polarization of the strong background breaks the isotropy of space.
This gives rise to the different refractive indices for parallel and
perpendicular polarizations of the probe in relation to the background
\cite{dittrich2003probing,Greger1}
\begin{equation}\label{RefIndizes}
n_\parallel=1+\frac{8\alpha}{45\pi}\frac{E^2}{E_{cr}^2},\qquad n_\perp=1+\frac{14\alpha}{45\pi}\frac{E^2}{E_{cr}^2}.
\end{equation}
The refractive indices in \eqref{RefIndizes} generate a difference in
the optical path length for parallel and perpendicular polarizations,
which is the cause of the flipping. The refractive indices
\eqref{RefIndizes} have been derived making use of 4-photon
interaction only, i.e., all but the first term
in \eqref{approxLHE} have been neglected.

The typical proposal for a setup for measuring flipping consists of
two coaxially \citep{Gies1} or nearly coaxially
\cite{luiten2004detection,Heinzl2006observeOffAxis} propagating
pulses, one of the two is a strong infrared (IR) background pulse and the
other a weak counter-propagating hard x-ray probe. A depiction of this
setup is shown in figure \ref{Fig:BireSetup}.

There are numerous analytical results of polarization flipping in the
literature. In the present paper the work published in
\cite{Gies1,Greger1} is used for benchmarking the efficiency and
accuracy of the algorithm. In particular, the flipping
probabilities for the settings in \citep{Gies1} and the
parametric dependencies derived in \citep{Greger1} are reproduced.

\section{Numerical results \label{Sec:Results}}
\subsection{High harmonic generation}
\begin{figure}[h]
\centering
\includegraphics[width=0.7\textwidth]{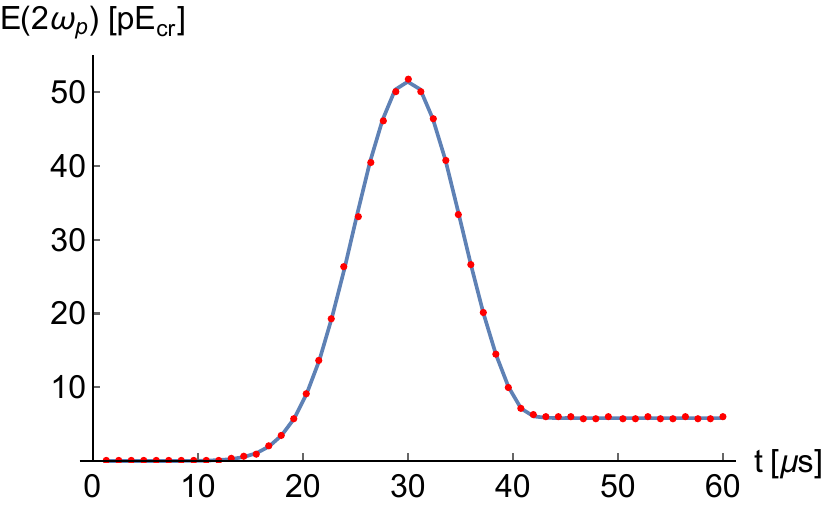}
\caption{Plot of the amplitude of the 2nd harmornic of the electric
  field versus time. The dots represent the simulation data while the
  solid line is the analyic approximation derived in
  \citep{Boehl1}. The error for all simulated data points lies below
  $3\%$\label{PatrickBenchHH}. At $t=60 \mu$s the probe has traversed
  the background implying that the signal represents the asymptotic
  2nd harmonic.}
\end{figure}

As can be seen in figure \ref{PatrickBenchHH} there is good agreement
between the analytic approximation and our simulation results of 2nd
harmonic generation. The relative error between numerical and
analytical results is less than $3\%$. While this error is worse than in
our numerical calculations of polarization flipping (see below) it has
to be noted, that the effect of high harmonic generation is of much
smaller relative magnitude and thus suffers more from errors
in numerical calculations.
Furthermore, a simulation with $\omega_{b} \neq 0$ allows to visualize
the $2\omega_{p} \pm \omega_{b}$ signal in the overlap field, which
comes from the 4-photon interaction and the $2\omega_{p}$ signal in
the asymptotic field at $t>50 \, \mu$s. The latter is generated by the 6-photon
contribution. This shows that the algorithm allows to time resolve the
different processes.

\subsection{Polarisation flipping: Experimental expectations}\label{subsec:PolFlip-Exp}
\begin{figure}[h]
\centering
\includegraphics[width=0.7\textwidth]{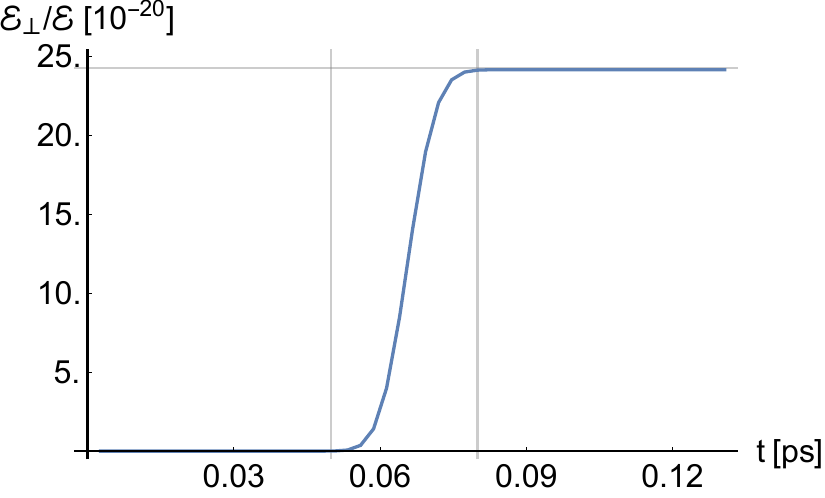}
\caption{Time evolution of the polarisation flipping for the
  parameters presented in Tab.\ref{Tab:GiesPulsParameter} with
  increased wavelength for the probe pulse ($\lambda_{p}=400$
  nm). The distance between the vertical lines corresponds to the
  interaction time $t_i$, within which 96\% of the polarisation flip
  occurs. The horizontal line notes the asymptotic relative flipping
  energy. \label{Fig:FlipTimeEvol}}
\end{figure}

For case (a) defined in \cite{Gies1}  the pulse parameters are given
in table \ref{Tab:GiesPulsParameter}.
\begin{table}[h!]
\centering
\begin{footnotesize}
\begin{tabularx}{\textwidth}{lll}  
\toprule
Sim. Box    & Size 					& $80 \mu$m \\
\midrule
Pump Pulse  & $|\vec{E}|_{\text{max}}$& $0.34$ m$E_{cr}$      \\
	        & $\hat{k} $			& $(-1,0,0)$      \\
	        & $x_0 $				& $60\,\mu$mm      \\
	        & $\lambda $			& $800\,$nm      \\
	        & $f $  				& $1.54\,$eV      \\
	        & $\Phi_t $				& $30\,$fs      \\
\midrule	        
Probe Pulse & $|\vec{E}|_{\text{max}}$& $0.05$ m$E_{cr}$      \\
	        & $\hat{k} $			& $(1,0,0)$      \\
	        & $x_0 $				& $20\,\mu$mm      \\
	        & $\lambda $			& $96\,$pm      \\
	        & $f $  				& $12.9\,$keV      \\
	        & $\Phi_t $				& $30\,$fs      \\
\bottomrule	        
\end{tabularx}
\end{footnotesize}
\caption{Parameters for probe and background beams presented in
  \cite{Gies1} \label{Tab:GiesPulsParameter}. We note that the background
  and probe pulses are coaxially counter-propagating. Furthermore, the
  background and probe pulse polarization are at an angle of $\pi/4$
  to each other and the peak to peak distance between the pulses at
  $t=0$ is $40\,\mu$m.}
\end{table}
It is worth noting that there are a couple of considerations to make when
simulating the setup presented in \cite{Gies1}. Firstly, since a 1D simulation
is used for case (a) in \cite{Gies1} in the present paper there is no lateral
dispersion of the background field. In order to account for the lateral
dispersion assumed in \cite{Gies1} the field strength in our
simulation needs to be reduced to the average
field intensity over the interaction time $t_i$ (see
figure \ref{Fig:FlipTimeEvol}).
Secondly, since in \cite{Gies1} the probe pulse is an x-ray pulse a
large number of grid points, $N_{\text{Point}}\approx 10^7$, is required to
remain below the Nyquist frequency, i.e., to avoid numerical artifacts due
to the undersampling of the oscillations. In order to avoid running such
large and expensive simulations a number of simulations with larger
wavelengths for the probe pulse (50 nm, 100 nm, 200 nm, 400 nm, 600
nm) are run and the flipping values are then extrapolated from these
results to x-ray wavelength.

Finally, as em-fields are being simulated instead of single photons
the only accessible data are the field values at any given time. To
compare these results to the ones in paper \cite{Gies1} the energies
in each polarization direction are computed as
\begin{equation}
\mathcal{E}_\perp=\sum_{x_i\in \mathcal{B}}\left(\vec{E}(x_i)\cdot
  \vec{\varepsilon}_\perp\right)^2,\quad\mathcal{E}_\parallel=\sum_{x_i\in
  \mathcal{B}}\left(\vec{E}(x_i)\cdot
  \vec{\varepsilon}_\parallel\right)^2,\quad
\mathcal{E}_{tot}=\mathcal{E}_\perp+\mathcal{E}_\parallel \, . \label{EnergyComputation}
\end{equation}
It can easily be checked, that
\begin{equation}
\frac{N_\perp}{N}=\frac{\hbar\omega N_\perp}{\hbar\omega
  N}=\frac{\mathcal{E}_\perp}{\mathcal{E}_{tot}} \, ,
\end{equation}
where $N$ and $N_\perp$ are the number of total and perpendicular
photons contained in the probe beam.

\begin{figure}[h]
\centering
\includegraphics[width=0.7\textwidth]{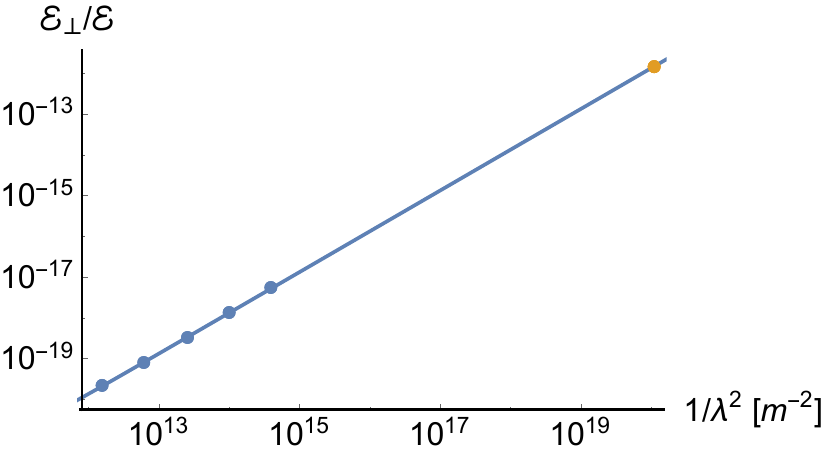}
\caption{Extrapolation of the wavelength in dependence of the
  polarisation flipping energy. The blue dots represent the relative
  polarisation energy at 800 nm, 400 nm, 200 nm, 100 nm and 50 nm from
  left to right. The solid blue line is the linear interpolation of
  these points with $1/\lambda^2_p$ as the linear variable. The orange
  point is the value given in \cite{Gies1}.\label{Fig:FlipFreqEvol}}
\end{figure}

Running the simulations using the algorithm presented in section
\ref{Sec:Numerical} and extrapolating the results assuming a
$1/\lambda^2$ scaling yields $N_\perp/N=1.42\cdot10^{-12}$ (see
figure \ref{Fig:FlipFreqEvol}), which represents a deviation of less than
2\% from the value obtained by Karbstein
et. al. ($N_\perp/N=1.39\cdot10^{-12}$) in \cite{Gies1}.
\subsection{Polarisation flipping: Parametrical dependencies}\label{subsec:PolFlip-Parameters}
For coaxially counter-propagating background and probe pulses the
polarisation flipping probability is given in the low energy
approximation $p_{b}p_{p}\ll m^2$, where $p_{b}$ is the momentum of
the background and $p_{p}$ the one of the probe pulse, by \citep{Greger1}
\begin{equation}
\mathbf{P}_{\text{flip}}=\frac{\alpha^2}{225}\frac{1}{\lambda_{p}^2}\sin^2(2\sigma)\left(\int
  \text{d}x^+\,\vec{E}\left(x^+\right)^2\right)^2 \, ,\label{flipProb}
\end{equation}
where $\sigma$ is the angle between the probe and background
polarizations. The main consequence of \eqref{flipProb} is the fact
that the pulse flipping probability depends solely on the background
pulse energy for any pulse shape. Besides that it states that the
flipping probability only depends on $\sigma$ and $\lambda_{p}$.

\begin{figure}[H]
\centering
\includegraphics[width=\textwidth]{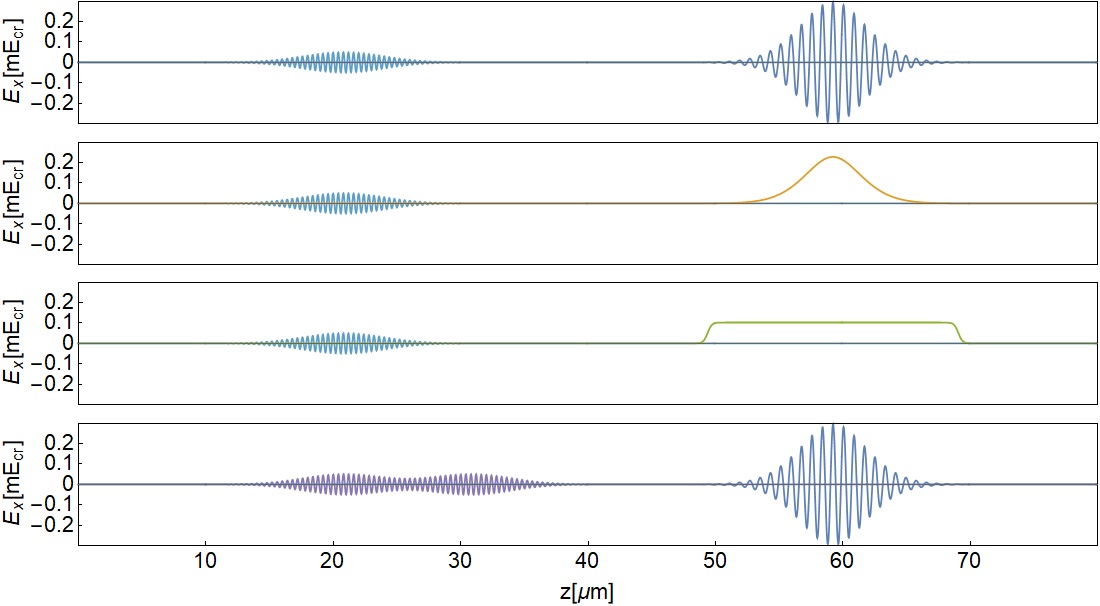}
\caption{ \label{Fig:BackgroundsBire} Different initial settings for
  the electric fields simulated. The parameters $\sigma$, $\lambda_{p}$,
  and $\mathcal{E}_{b}$ are the same for all the settings. The reference
  setting (blue, top) is the one used for the reproduction of the
  results given in \citep{Gies1} with $\lambda_{p}=400$ nm, see
  \ref{subsec:PolFlip-Exp}. Then, there are the 0-frequency background
  (orange, 2nd from top), the shock regime background (green, 3rd from
  top), and the chirped probe (purple, bottom). The
  color scheme corresponds to the one used in figure 
  \ref{Fig:EnergiesBackgroundsBire}.}
\end{figure}

\begin{figure}[H]
\centering
\includegraphics[width=0.6\textwidth]{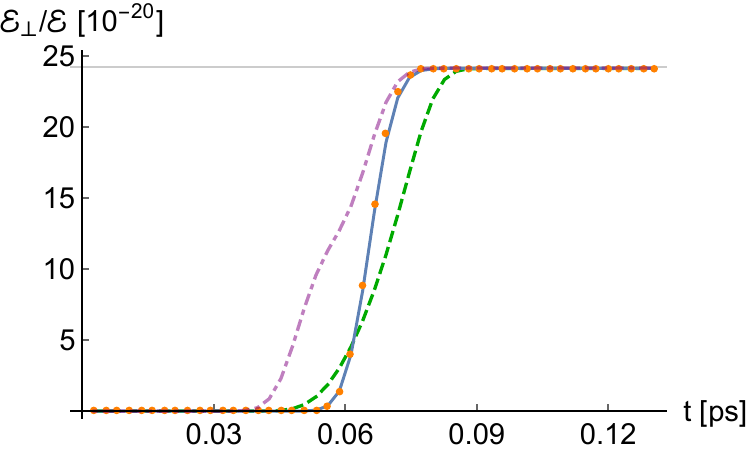}
\caption{Plot of the relative flipping amplitude versus time for the
  different backgrounds shown in figure \ref{Fig:BackgroundsBire}. The
  color scheme is the same as in figure \ref{Fig:BackgroundsBire}.
  The relative numerical error for the asymptotic value (horizontal
  line) computed with the help of relation \eqref{flipProb} is less or equal than
  $0.5\%$ for all the cases. \label{Fig:EnergiesBackgroundsBire}}
\end{figure}

\begin{figure}[H]
\begin{subfigure}{.48\textwidth}
  \centering
  \includegraphics[width=\linewidth]{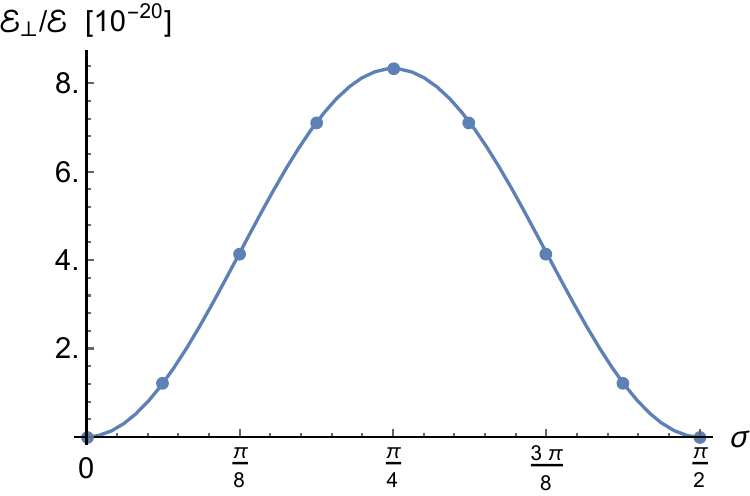}
  \label{Fig:FlipSigEvol}
\end{subfigure}%
\begin{subfigure}{.52\textwidth}
  \centering
  \includegraphics[width=\linewidth]{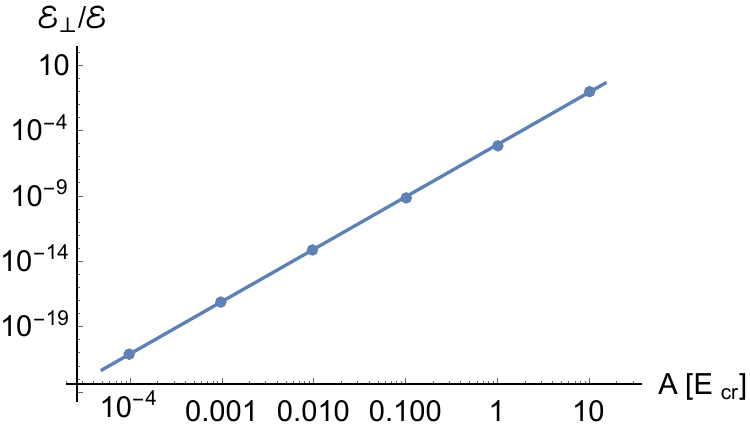}
  \label{Fig:FlipEneEvol}
\end{subfigure}
\caption{Left: Plot of the relative polarization flipping vs. angle $\sigma$
  between probe and background pulse polarizations using the settings
  from table \ref{Tab:GiesPulsParameter} with $\lambda_{p}$ set to $400$
  nm. The dots are the values obtained via simulation and the solid
  line is the prediction using \eqref{flipProb}. Note that the
  relative error is less or equal to $0.3\%$ for all data points. Right:
  Plot of relative polarization flipping vs. background pulse
  amplitude $A$ using the parameters from table
  \ref{Tab:GiesPulsParameter} with $\lambda_{p}$ set to $400$
  nm. The dots are the values gained with the help of simulations and
  the solid line is the prediction using \eqref{flipProb}. The
  relative error is less or equal to $0.3\%$ for all data points.}
\label{Fig:FlipEvol}
\end{figure}

In order to verify that the probability for flipping depends only on the
pulse energy a set of different background and probe profiles with the 
same energy, $\sigma$, and $\lambda_{p}$ (see
figure \ref{Fig:BackgroundsBire}) is chosen.
As figure \ref{Fig:EnergiesBackgroundsBire} shows the statement holds for
the asymptotic field, but the form of the polarization flip during the interaction
and the duration of the interaction vary heavily.
The reproduction of the $\lambda_{p}$ scaling by the algorithm can
be seen in figure \ref{Fig:FlipFreqEvol}. The $\sigma$ and
energy scalings remain to be checked. Different values for the peak
field strength $A$ and relative polarization angle $\sigma$ starting
with the parameters given in table \ref{Tab:GiesPulsParameter} and
$\lambda_{p}=400$ nm are simulated. The results are shown in figure
\ref{Fig:FlipEvol}.
Besides the small error between numerical and analytical results of
$0.3\%$ in all simulations the flipping probability is
$1.2 \cdot 10^{-27}$ at $\sigma=\pi/2$, which is larger than the
numerical noise of $\sim 4\cdot 10^{-33}$. The signal of the flipping
probability at $\pi/2$ should not be there according to
\eqref{flipProb}. It is due to high harmonic generation neglected 
in the analytical derivation of the flipping probability in
\eqref{flipProb}. Furthermore, extreme testing of the numerical
algorithm shows, that it remains stable up to $A=100\, E_{cr}$, which
is well beyond the limits of the weak field
expansion.

\subsection{Parallelization scaling}
Measurements of weak scaling show a $1.09 \cdot\#_{cores}$ dependence
on the total core time. Similarly, a factor of $1.04$ is measured for
strong scaling.

\section{Results in higher dimensions}
\begin{figure}[p]
\includegraphics[width=0.9\textwidth]{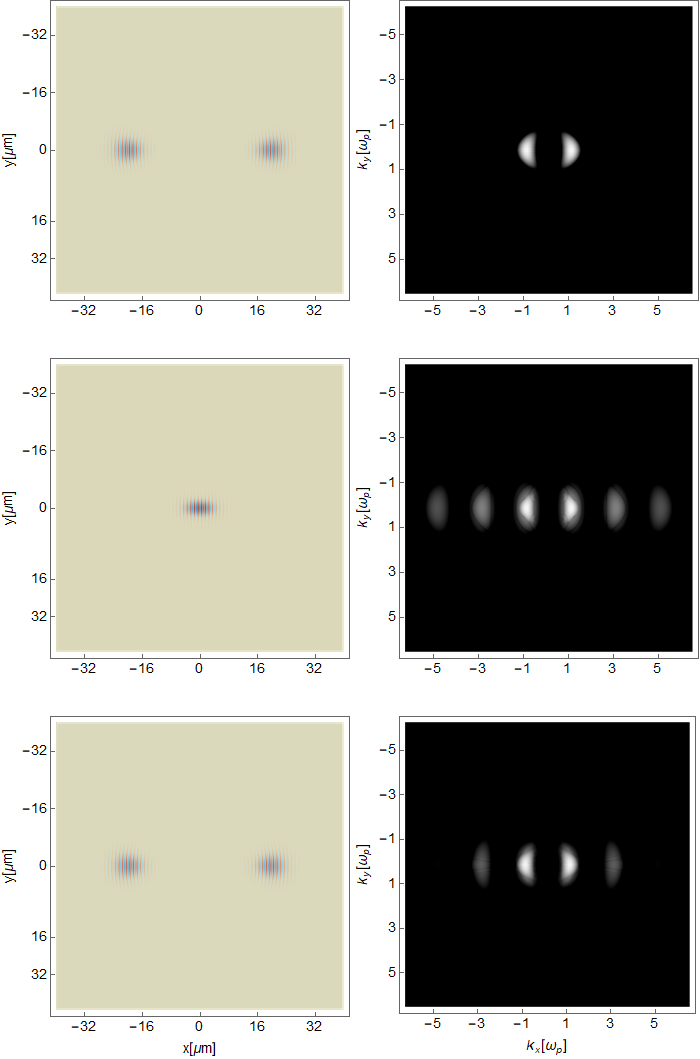}
\caption{Plots of simulated values for $E_z$ for two coaxially
  propagating Gaussian pulses with $E_0=0.05 E_{cr}$. The plots to the left
  are in physical space and the ones to the right are in frequency
  space. The two in the  top are the initial settings, the ones in the
  middle show the overlap state, and the ones in the bottom represent
  the field configuration after the collision.\label{Fig:2DCoaxial}}
\end{figure}

Having verified the validity of the algorithm the next step is to
simulate higher dimensions. The first simulation in higher dimensions
is the collision of two coaxial Gaussian pulses
in 2D (see figure \ref{Fig:2DCoaxial}). As in the 1D case the generation of
higher harmonics is observed due to the inclusion of the 6-photon
scattering diagram in the numerical simulation. Furthermore, the
sharpening of the pulses as predicted by \citep{Gies1} is observed
in the asymptotic field.

\begin{figure}[p]
\includegraphics[width=0.9\textwidth]{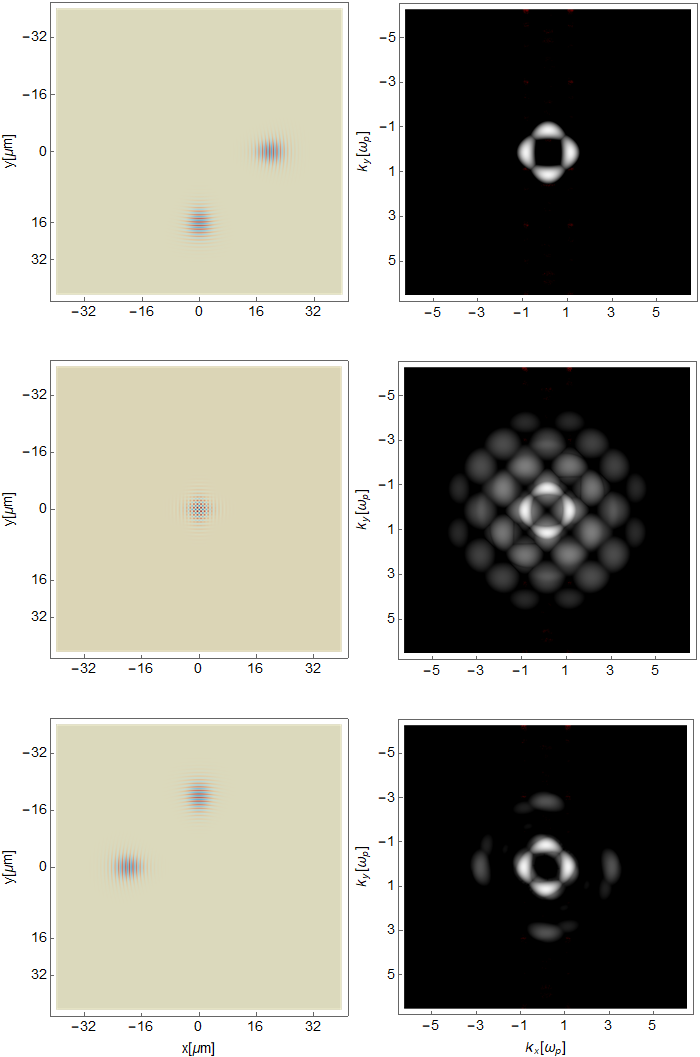}
\caption{Plots of simulated values for $E_z$ for two orhogonally
  propagating Gaussian pulses with $E_0=0.05 E_{cr}$. The plots to the left
  are in physical space and the ones to the right are in frequency
  space. The two in the top are the initial settings, the ones in the
  middle show the overlap state, and the ones in the bottom represent
  the field configuration after the
  collision.\label{Fig:2DOrthogonal}}
\end{figure}

A feature-rich simulation is shown in figure
\ref{Fig:2DOrthogonal}. The pulses in the figure are no longer
propagating along the same axis. As a result a myriad of mixing
processes caused by the 4-photon interaction opens up. Once again, it
can be observed how theses signatures vanish in the asymptotic
fields. It is, however, worth noting that contrary to the 1D case the
asymptotic $3\omega_p$ signal is slightly of axis causing it to split
of from the main pulse. 

Furthermore, it is verified via turning off the 4- and 6-photon
scattering diagrams separately, that the higher harmonic signal in the
asymptotic field is once again solely generated by the 6-photon
scattering while the 4-photon contribution does not generate anything
beyond scattering signals in the $\omega=\omega_{p}$ spectral
region. Note that in the orthogonal case the signal is not symmetric
along the $\omega_x=0$ and $\omega_y=0$ axes. However, symmetry is
conserved along the symmetry axes $\omega_x + \omega_y=0$, which
corresponds to the initial symmetry of the setup.

It has to be mentioned that the symmetry of the simulated fields has
to be conserved by a accurate and consistent numerical algorithm. This
statement is of the utmost importance as it would be clearly
non-physical to violate basic symmetries of the underlying equations.
Symmetry violation is a clear sign that an algorithm is not working
properly.
\begin{table}[h!]
\centering
\footnotesize
\begin{tabularx}{\textwidth}{lll}  
\toprule
Sim. Box    & Size 					& $80 \mu$m$\times80 \mu$m \\
\midrule
Background Pulse  & $\vec{E}_0$			& $(0,0,50)$ m$E_{cr}$      \\
	        & $\hat{k} $			& $(1,0,0)$      \\
	        & $\vec{x}_0 $			& $(40,40,0)\,\mu$mm      \\
	        & $w_0 $				& $4.6\,\mu$m      \\
	        & $z_r $				& $16.619\,\mu$m      \\
	        & $\lambda $			& $4\,\mu$m      \\
	        & $z_t $				& $20\,\mu$m      \\
	        & $\Phi_t $				& $4.5\,\mu$m      \\
\midrule	        
Probe Pulse & $\vec{E}_0$			& $(0,0,50)$ m$E_{cr}$      \\
	        & $\hat{k} $			& $(-1,0,0)$     \\
	        & *** &  {\it all other parameters as for the background}      \\
\bottomrule	        
\end{tabularx}
\caption{\label{Tab:2DCoaxial} Parameters for the collision of two pulses
  with the same frequency in 2 dimensions. See figure 
  \ref{Fig:2DCoaxial} for results. Note that they are also the same
  for the orthogonal case except for the $\hat{k}$, which
  becomes $(0,1,0)$ }
\end{table}


\section{Discussion and conclusion\label{Sec:Discussion}}
A numerical solver for the nonlinear Heisenberg-Euler wave equations
has been derived. Only two approximations have been made: (i) The
wavelengths involved are larger than the Compton wavelength and (ii)
the involved fields are weaker than the critical field strength
$E_{cr}$. Furthermore, we have shown that the numerical results 
agree with a range of analytic results derived in the literature
\cite{Boehl1,Boehl2,Gies1,Greger1}.

It is astonishing that there is excellent agreement between the
simulation results and the analytic benchmarks, since in \cite{Gies1} and in
\cite{Greger1} the x-ray probes are considered to be a point like photons,
whereas here it they are considered to be a counter-propagating electromagnetic
pulses. However, as seen in the discussion of the birefringence
results in the present paper the flipping probability is mostly
independent of pulse shape. It depends only on the frequency of the
pulse and the relative angle between the polarizations of probe and
background pulses. Due to this property of the interaction it is clear
that the analytic calculations \cite{Gies1,Greger1} show such good agreement with the
numerical approach in this paper.

Furthermore, while all analytical calculations in the literature consider only
one effect at a time the simulations in this paper reproduce all of
the effects simultaneously. Under certain conditions, a computation
for a single effect is faster using the aforementioned analytical
calculations. However, the simulated data represents the complete
picture without having to consider how different effects interfere
with each other. In addition, no restrictive assumptions need to be
made about the initial setup. As a consequence numerical simulations can
be used for real world experimental settings.

The properties of the algorithm in higher dimensions remain to be
tested against predictions made for example in \citep{Greger2}. They
will require a large number of expensive simulations to scan all the
interesting parameters. Due to the fact that no specific assumptions
concerning spatial directions have been made during the derivation of
our numerical method we expect the algorithm to work in 3D.

Finally, as the number of assumptions that have been made in the
course of the derivation of the numerical algorithm is small it is
possible to use the latter to simulate all the predicted light by
light scattering effects such as focusing or diffraction-angle
specific polarization for a wide range of initial settings. This in
turn may be used to develop tomographic methods for strong pulse
characterization.

\section{Outlook \label{Sec:Outlook}}
The ability to time resolve the nonlinear vacuum optical processes
opens an array of interesting new investigations. For example, the
intensity of the background pulse can be increased up to the point
where the low energy approximation used to derive \eqref{flipProb} no
longer holds and the relative orthogonal polarization energy displays
Rabi like oscillations \cite{griffiths1999electrodynamics}.
\begin{figure}[h]
\centering
\includegraphics[width=0.55\textwidth]{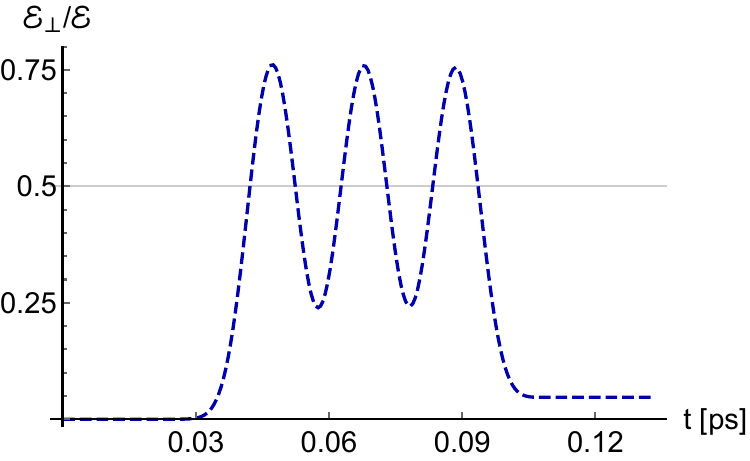}
\caption{Plot of the time evolution of the relative perpendicular
  polarisation of the probe vs. time. The initial setup corresponds to
  the one shown in figure \ref{Fig:BackgroundsBire}[green line] with
  the exception of the amplitude of the background pulse being
  $20\,E_{cr}$. The fact that the amplitude of the oscilations seen in
  the figure remains below 1 is due to the finite pulse width of the
  probe, which causes its different parts to have propagated through
  different amounts of the background at any given time.\label{OutlookOsc} }
\end{figure}
An example of this effect can be seen in figure \ref{OutlookOsc},
where the intensity of the background has been increased to the ultra
high field regime ($20\,E_{cr}$). The latter is only meant as a proof
of the existence of Rabi-like oscillations under extreme conditions. However,
the ultra-high field regime is at the limits of the validity range of
the weak-field approximation of the HE-Lagrangian \eqref{approxLHE}. It is
also well past the point, where pair production \eqref{Eq:PairProd} may
be safely neglected. However, the effect is still worth considering as
it may also be triggered by a longer weaker pulse.

\section*{Aknowledgments}
 A.P.D. would like to acknowledge the support of P. B\"ohl during the
familiarization period with the matter. This work was funded by the
Munich Cluster of Excellence (MAP), by the international Max-Planck
Research School for Advanced Photonic Sciences (IMPRS-APS), and the
Transregio TR-18 project B12.

\section*{References}
\bibliography{PaperBib}

\end{document}